\documentclass[a4paper,12pt]{article}
\usepackage[T1]{fontenc}
\usepackage[utf8]{inputenc}
\usepackage[affil-it]{authblk}
\usepackage{slashed}
\usepackage{amssymb, amsmath}
\usepackage{mathtools}
\usepackage{stmaryrd}
\usepackage{xcolor}
\usepackage{a4wide}
\usepackage{bbding}
\usepackage{newtxtext, newtxmath}
\usepackage{mathrsfs} 
\usepackage{bbm}
\usepackage{hyperref}
\hypersetup{
    colorlinks,%
    citecolor=green!50!black,%
    filecolor=black,%
    linkcolor=blue,%
    urlcolor=blue
}
\usepackage[numbers,sort&compress,merge]{natbib}
\usepackage{graphicx}
\usepackage{filecontents}

\usepackage{graphicx}
\usepackage[toc,page]{appendix}

\newtheorem{theorem}{Theorem}[section]
\newtheorem{proposition}{Proposition}[section]

\newcommand{\scalarProd}[2]{ \left\langle {#1} , {#2} \right\rangle } 
\newcommand{\scalarProdDirac}[2]{ \left\langle {#1} | {#2} \right\rangle } 
\newcommand{\norm}[1]{\left\| {#1} \right\|}

\newcommand{\sA}{\mathscr{A}}

\newcommand{\sB}{\mathscr{B}}

\newcommand{\sC}{\mathscr{C}}
 
\newcommand{\C}{\mathbb{C}}
\newcommand{\CAR}{\mathscr{C\!A\!R} (\mathfrak{K})}

\newcommand{\sD}{\mathscr{D}}
\newcommand{\rd}{\,\mathrm{d}}

\newcommand{\sF}{\mathscr{F}_{\Psi}}
\newcommand{\kF}{\mathfrak{F}_{\Psi}}
\newcommand{\bkF}{\boldsymbol{\mathfrak{F}}_{\Psi}}

\newcommand{\kH}{\mathfrak{H}_{\Phi}}
\newcommand{\bkH}{\boldsymbol{\mathfrak{H}}_{\Phi}}

\newcommand{\kK}{\mathfrak{K}}

\newcommand{\sM}{\mathscr{M}}

\newcommand{\sR}{\mathscr{R}_{\Phi}}

\newcommand{\kS}{\mathfrak{S}}
\newcommand{\ks}{\mathfrak{s}}

\newcommand{\sV}{\mathscr{V}}
\newcommand{\sW}{\mathscr{W}}

\newcommand{\LTwoCauchy}{L^{2} (\mathscr{S\!M} \upharpoonright \mathscr{C})}

\newcommand{\R}{\mathbb{R}}

\newcommand{\ri}{ \mathrm{i} }

\newcommand{\supp}[1]{ \mathrm{supp} ({#1}) }

\newcommand{\fbracket}[1]{ \left( {#1} \right) } 
\newcommand{\sbracket}[1]{ \left\{ {#1} \right\} }

\newcommand{\dist}{\mathrm{dist}}

\newcommand{\PA}{P_{\mathscr{A}}}
\newcommand{\PB}{P_{\mathscr{B}}}
\newcommand{\PBprime}{P_{\mathscr{B}'}}
\newcommand{\PV}{P_{\mathscr{V}}}

\newcommand{\SigmaPsi}{\Sigma_{\Psi}}

\newcommand{\SigmaPsiBprime}{\Sigma_{ \Psi, \mathscr{B}'}}

\newcommand{\SigmabetaBprime}{	\Sigma_{ \beta, \scriptscriptstyle{B'} }	}

\newcommand{\csch}{\mathrm{csch}}

\newcommand{\cst}{\mathrm{cst}. \,}

\title{Relative entanglement entropy of thermal states of 
Klein-Gordon and Dirac quantum field theories}
\author{Onirban Islam\thanks{mmoi@leeds.ac.uk}}
\affil{School of Mathematics, University of Leeds, LS2 9JT, UK}
\date{23 February 2020}

\begin{document}
\maketitle
\begin{abstract}
	An upper bound of the relative entanglement entropy of thermal states at an inverse temperature $\beta$ of linear, massive Klein-Gordon and Dirac quantum field theories across two regions, separated by a nonzero distance $d$ in a Cauchy hypersurface of an ultrastatic spacetime has been computed. 
	This entanglement measure is bounded by a negative constant times $\ln | \tanh (\pi d/ 2 \beta) |$ which signifies power law decay for asymptotic $d$ where the exponent depends on $\beta < \infty$.  
\end{abstract}

\section{Introduction} 
The relative entanglement entropy is a good measure of entanglement 
for generic states in quantum field theory in Minkowski and curves spacetimes~\cite{Hollands_Sanders_Springer_2018}. 
Hitherto, upper bounds of this entanglement measure have been computed for 
pure global states~\cite{Hollands_Sanders_Springer_2018, Hollands_JMP_2018, Takayanagi_arXiv_2018}, 
albeit the measure is well applicable to global mixed states. 
Thermal states are a class of global mixed states characterized by the 
Kubo~\cite{Kubo_JPSJapan_1957}-Martin-Schwinger~\cite{Martin_Schwinger_PR_1959} 
condition
(the mathematical rigour formulation is due to Haag \textit{et al}.~\cite{Haag_CMP_1967}; 
see also the 
reviews~\cite{Kay_Wald_PR_1991, Sanders_IJMPA_2013} 
and the 
monograph~\cite{Bratteli_Springer_1997} 
for details),  
that describe a wide range of phenomena in elementary particle physics, 
cosmology and condensed matter physics.

In this report, we have considered 
the Klein-Gordon and the Dirac quantum field theories 
in a Cauchy hypersurface of a ultrastatic spacetime. 
By making use of the techniques developed in~\cite{Lechner_Sanders_Axioms_2016, Hollands_Sanders_Springer_2018, Hollands_JMP_2018},  
we implement the required computational modifications to account the mixed nature of thermal states to compute the relative entanglement entropy across two regions of non-vanishing distance $d$ in the Cauchy hypersurface. 
These modifications result a different bound for the entanglement measure of thermal state compared to the ground state. 
More precisely, the relative entanglement entropy is bounded by 
$ - \cst \ln | \tanh (\pi d/ 2 \beta) |$ for both theories which implies that this exhibits power law decay with asymptotic $d$, where the exponent depends on the inverse temperature $\beta < \infty$.

Our sign convention for the spacetime metric is mostly minus and 
the scalar curvature $R$ is positive on the sphere. 
At first, the Dirac field is presented inasmuch 
that requires more general strategy than the Klein-Gordon field 
to prove the results.

\section{Relative entanglement entropy of Dirac-Majorana QFT}
\label{sec: REE_Dirac}
We begin the section by a minimal review of the quantization of a classical, linear Dirac-Majorana field in a $D \in \{4, 8, 9, 10\}~\mathrm{mod}~8$-dimensional\footnote{The 
restriction of the spacetime dimensions stems from the existence of the time-reversal operator~\cite[Lem. V.1]{Hollands_JMP_2018} required to prove our results and existence of (pseudo-)Majorana spinors~\cite{Kugo_Townsend_NuclPhysB_1983, Trautman_JGP_2008}.} 
simply-connected ultrastatic spin-spacetime 
$\sM = \R \times \sC$ with metric $g = \mathrm{d} t^{2} - h$ and $H^{1} (\sM) = 0$,
where $h$ is the Riemannian metric, independent of $t$ on the on the Riemannian spin-manifold $(\sC, h)$ which is assumed to be complete. 
We assume that $e^{0}$ be the forward directed time-like normal vector field on the Cauchy hypersurface $\sC$ and let $\kK := (\LTwoCauchy, \scalarProd{\cdot}{\cdot}, \Gamma)$ be the space $\LTwoCauchy$ of square-integrable Cauchy data $k$  with respect to the positive definite, non-degenerate hermitian inner product $\scalarProd{\cdot}{\cdot}$, equipped with an antilinear involution $\Gamma$~\cite{Trautman_JGP_2008}. 
Here $\mathscr{S\!M} \to \sM$ is the spinor bundle whose details is not important for the primary content of this article, and so we resist its exposition and refer~\cite{Antoni_Hollands_CMP_2006, Baer_Ginoux_Springer_2012} and references therein, for instance.

We will quantize the classical spinors in the selfdual framework by Araki~\cite{Araki_RIMS_1970} 
(see also~\cite{Antoni_Hollands_CMP_2006, Baer_Ginoux_Springer_2012}). 
In this approach a quantum Fermi field $\psi$ defined by a Cauchy hypersurface $\sC$ is an algebra of canonical anticommutation relations $\CAR$-valued $\C$-linear distribution $\psi : \kK \to \CAR$. 
The algebra $\CAR$ is defined as the free unital *-algebra over $\kK$ generated by the symbols $\mathbbm{1}, \psi (k), \psi (l)^{*}$ modulo the relations
\begin{equation}
    \psi (k)^{*} = \psi (\Gamma k),  
    \quad   
    \sbracket{ \psi (k), \psi (l)^{*} } = \scalarProd{l}{k} \mathbbm{1}, 
    \quad \forall k, l \in \kK. 
\end{equation}
There exists a unique $C^*$-norm~\cite{Araki_RIMS_1970, Baer_Ginoux_Springer_2012} on $\CAR$ whose closure $\overline{\CAR}$ defines a $C^{*}$-algebra.   
The local algebra of field $\overline{\CAR}$ corresponding to some bounded open region $\sV \subset \sC$, is by definition, 
the $C^{*}$-subalgebra generated by all elements of the form $\psi (k)$ with $\supp{k} \subset \sV$. 
The local algebra of observables $\sD (\sV)$ is the $C^{*}$-subalgebra of $\overline{\CAR}$, consisting of only the even elements of 
$\overline{\CAR}$.

An algebraic thermal state $\Psi$ is a linear functional of $\sD(\sC)$, uniquely characterized by the bounded operator~\cite[Prop. 1]{Balslev_CMP_1968},~\cite[Thm. 3]{Araki_RIMS_1970} 
(see also, e.g.,~\cite{Bratteli_Springer_1997})
\begin{equation} \label{eq: thermal_bounded_CAR}
    \SigmaPsi := \tanh{\frac{\beta H}{2}}, 
\end{equation}
where $H$ is the Dirac Hamiltonian and $\beta$ is the inverse temperature.
Applying the Gelfand-Naimark-Segal (GNS) construction for $\Psi$, we obtain the GNS-triple $(\pi_{\Psi}, \bkF, \Omega_{\Psi})$ where $\Omega_{\Psi} \in \C$ is the GNS-vector and $\pi_{\Psi}$ is a $*$-representation of $\sD (\sC)$ on the GNS-Hilbert space $\bkF := \C \oplus ( \oplus_{ n = 1 }^{\infty} \bigwedge^{n} \kF )$, 
where the  ``1-particle'' Hilbert space $\kF$ is constructed from $\kK$ by dividing out $ \ker (I + \SigmaPsi) / 2$, and taking closure with respect to the inner product~\cite{Antoni_Hollands_CMP_2006}
\begin{equation}
    ([k] | [l])_{\Psi} := \frac{1}{2} \scalarProd{k}{ (I + \SigmaPsi) l }, 
    \quad \forall k, l \in \kK. 
\end{equation}
Employing the GNS-representation, we define the local von Neumann algebra of observables as
\begin{equation}
    \sF (\sV) := \pi_{\Psi} ( \sD (\sV) )'',
\end{equation} 
where $''$ is the double commutant.

Suppose that $\sA, \sB \subset \sC$ be any two open sets. 
We recall that 
(see, e.g.,~\cite{Hollands_Sanders_Springer_2018} 
and references therein for the required background of entanglement in 
quantum field theoretic setting,  
and 
e.g.~\cite{Bratteli_Springer_1997, Kadison_Ringrose_II_AMS_1997} 
for the relevant concepts of von Neumann algebra) 
the relative entropy $S (\omega, \rho)$ between 
two global states $\omega$ and $\rho$ is defined\footnote{The 
explicit formulation is not required for the computation in this article.} 
by the Araki's formula~\cite{Araki_RIMS_1976, Araki_RIMS_1977}, and the relative entanglement entropy $E(\omega)$ of a global state $\omega$ 
across the bipartite system $(\sF (\sA), \sF (\sB))$ of two commutating local von Neumann algebras (in the standard form) $\sF (\sA)$ and $\sF (\sB)$ is defined by the infimum of $S (\omega, \rho)$ over the weak $*$-convex hull of separable states $\rho$ 
for all type I von Neumann factors that splits the bipartition~\cite{Narnhofer_RepMathPhys_2002}.

Now, we state the main result of this section in 
\begin{theorem} \label{thm: REE_Dirac_thermal}
    Let $\sC$ be a static Cauchy hypersurface in a geodesically complete, simply connected, $D \in \{4,8,9,10\}~\mathrm{mod}~8$-dimensional ultrastatic spin-spacetime $(\sM = \R \times \sC, g = \mathrm{d} t^2 -h)$ such that $\inf (m^{2} + R (x) / 4) > 0$  holds on $(\sC, h)$,  
    where $R$ is the scalar curvature of $(\sC, h)$ and $m$ the mass of the Dirac field. 
    Then the relative entanglement entropy $E (\Psi)$ for a thermal state $\Psi$ at an inverse temperature $\beta$ of the linear Dirac quantum field theory    
    between any open subsets $\sA, \sB \subset \sC$ such that $d := \dist (\sA, \sB) \geq \delta > 0$, is bounded by
    \begin{equation}
	    E (\Psi) 
	    \leq 
	    - \cst \ln \left| \tanh \frac{\pi}{2} \frac{d}{\beta} \right|,  
    \end{equation}
    where $\cst$ is a positive constant which depends on $M := \sqrt{\inf (m^{2} + R (x) / 4)}, \delta$ and the geometry within a $\delta$-neighborhood of $\sA$,  
    and so is independent of $d$ and $\beta$.
\end{theorem}
\textbf{Proof} 
We employ the same strategy as in~\cite[Sec. V]{Hollands_JMP_2018} 
with the essential replacement of local von Neumann algebra 
and Fock space for ground state by those, 
$\sF (\sV)$ and $\bkF$  of the thermal state $\Psi$, respectively. 
Since, the GNS-vector $\Omega_{\Psi}$ is cyclic and separating~\cite[Thm. 4.8]{Strohmaier_CMP_2000} for $\sF (\sB')$  
where $\sB' := \sC \setminus \overline{\sB}$, 
we consider the pair $(\sF (\sB'), \Omega_{\Psi})$ 
and define the map 
\begin{equation}
    \Xi_{\Psi}^{\sA} : \sF (\sA) \to \bkF, \quad A \mapsto 
    \Xi_{\Psi}^{\sA} (A) := \Delta_{\Psi, \sB'}^{\frac{1}{4}} A \Omega_{\Psi}.
\end{equation}
As a consequence of Theorems 3 and 4 in~\cite{Hollands_Sanders_Springer_2018}: 
\begin{equation}
    E (\Psi) 
    \leq 
    \log \min 
    \fbracket{ \norm{\Xi_{\Psi}^{\sA}}_{1}, \norm{\Xi_{\Psi}^{\sB}}_{1} }, 
\end{equation}
where the minimum is taken to get a quantity that is symmetric under the 
exchange of $\sA$ with $\sB$.  
Thus we look for an upper bound of the $1$-nuclear norms 
(see, e.g.~\cite{Lechner_Sanders_Axioms_2016} 
for mathematical details)
of $\Xi_{\Psi}^{\sA}$ and $\Xi_{\Psi}^{\sB}$.   
We note that the $1$-particle Dirac Hamiltonian 
commutes with the time-reversal operator 
\cite[Lem. V.1]{Hollands_JMP_2018}, 
and so does the operator $\Sigma_{\Psi}$ by the spectral theorem.  
This implies that the $1$-particle Hilbert space $\kF$ is invariant under 
the time-reversal operator and 
so is its any standard real subspace 
(see, e.g.~\cite{Longo} for details on standard subspace),  
as the time-reversal operator preserves localization 
\cite[Lem. V.1]{Hollands_JMP_2018}.  
Furthermore, one can employ analogous arguments as in 
\cite[Lem. V.2]{Hollands_JMP_2018} 
to deduce that the $1$-particle modular operator 
commutes with the time reversal operator. 
Then the arguments (for instance, the doubling prescription, construction of closed complex linear subspaces of $\kF$) in~\cite[Sec. V]{Hollands_JMP_2018} smoothly flow over and we can apply Theorem 3.11 and Theorem 3.5 in~\cite{Lechner_Sanders_Axioms_2016} with Proposition IV.2 in~\cite{Hollands_JMP_2018} to deduce that the nuclear norm of $\Xi_{\Psi}^{\sA}$ is bounded by the trace-norm of $(P_{\sB'} - \Sigma_{\Psi, \sB'}^{2})^{1/4} \! \upharpoonright \! \kK (\sA)$ where $\kK (\sA) := \kK \! \upharpoonright \! \sA$, $P_{\sB'} : \kK \to \kK (\sB')$ is the projector and $\Sigma_{\Psi, \sB'} := P_{\sB'} \Sigma_{\Psi} P_{\sB'}$ is the restriction of $\Sigma_{\Psi}$ on $\kK (\sB')$ 
(cf.~\cite[Eq. 33]{Hollands_JMP_2018}). 
Altogether, we arrive at 
\begin{equation}
    E (\Psi) 
	\leq 
	\cst \norm{ (P_{\sB'} - \Sigma_{ \Psi, \sB' }^{2} )^{\frac{1}{4}} \! \upharpoonright \! \kK (\sA) }_{1}. 
\end{equation} 
Thus, our task boils down to the

\begin{proposition} 
    In the preceding notations, on a complete Riemannian spin-manifold $(\sC, h)$, 
    we have for any $\delta > 0$
    \begin{equation}
	    \norm{ (P_{\sB'} - \Sigma_{ \Psi, \sB' }^{2} )^{\frac{1}{4}} \! \upharpoonright \! \kK (\sA) }_{1} 
	    \leq 
	    - \cst \ln \left| \tanh \frac{\pi}{2} \frac{d}{\beta} \right|, 
	    \quad \forall d \geq \delta, 
    \end{equation} 
    where $\cst$ is a positive constant independent of $d$ and $\beta$ but may depend on $\delta$ and $M$.
\end{proposition} 
\textbf{Proof}: 
Complying to the Prop. V.1 in~\cite{Hollands_JMP_2018}, 
we note that $0 < \SigmaPsi^{2} < I$ in contrast to the pure states 
(where it is an involution). 
Let $\PV$ be the restriction map $\kK \to \kK (\sV)$ for $\sV = \sA, \sB, \sB'$ 
and compute using the properties of projectors  
\begin{eqnarray}
    \PA (\PBprime - \SigmaPsiBprime^{2}) \PA 
    & = & 
    \PA (I - \SigmaPsi \PBprime \SigmaPsi) \PA 
    \nonumber \\ 
    & \leq & 
    \PA (I - \SigmaPsi \PBprime \SigmaPsi) \PA + \PA \SigmaPsi^{2} \PA   
    \nonumber \\ 
    & = & 
    \PA + | \PB \SigmaPsi \PA |^{2} 
    \nonumber \\ 
    & \leq & 
    (\PA + | \PB \SigmaPsi \PA |)^{2}.  
    \label{eq: Dirac_1}
\end{eqnarray}
Employing the definition of projection operator and 
operator monotone property of the square root one can show that 
(cf.~\cite[Eqs. 37 and 38]{Hollands_JMP_2018})
\begin{equation}
	\left| ( \PBprime - \SigmaPsiBprime^{2} )^{ \frac{1}{4} } \PA \right| 
	\leq
	\sqrt{ \PA ( \PBprime - \SigmaPsiBprime^{2} ) \PA } 
    \leq 
    \check{X} \PA + | \PB ( I - \hat{X} ) \SigmaPsi \check{X} \PA |, 
\end{equation}
where we have used~\eqref{eq: Dirac_1} 
and $\check{X}, X, \hat{X}$ be the multiplication operators by 
the smooth functions $\check{\chi}, \chi, \hat{\chi}$, respectively. 
These smooth functions are defined as follows 
\cite{Hollands_JMP_2018}. 
We introduce two intermediate regions between $\sA$ and $\sB$, 
called $\check{\sV}$ and $\hat{\sV}$ such that 
$\sA \subset \check{\sV} \subset \hat{\sV} \subset \sB'$. 
Then the smooth functions $ \check{\chi}, \hat{\chi}, \chi $ are defined by: 
(i) $\supp{\check{\chi}} \subset \check{V}$ and $\check{\chi} \equiv 1$ on $\sA$;  
(ii) $\supp{\hat{\chi}} \subset \sB'$ and $\hat{\chi} \equiv 1$ on $\hat{\sV}$, $1- \hat{\chi} \equiv 1$ on $\sB$;  
(iii) The distance $\dist (\supp{\check{\chi}}, \supp{1-\hat{\chi}} ) = d - \varepsilon $, where $d :=\dist (\sA, \sB) $ and $ \varepsilon > 0 $ is thought of as small;   
(iv) $\chi$ is a function of compact support such that $\chi \equiv 1$ on $\supp{\check{\chi}}$, $\supp{\chi} \subset \supp{\hat{\chi}}$ and 
$ \dist( \supp{\chi}, \supp{1- \hat{\chi}} ) = d - 2 \varepsilon $ 
(see Fig. 1 in~\cite{Hollands_JMP_2018} for a schematic visualization). 

We utilize the supports of the smooth functions $\check{\chi}, \chi, \hat{\chi}$ 
and properties of the trace-norm and operator norm to deduce  
\begin{eqnarray}
	\norm{ ( \PBprime - \SigmabetaBprime^{2} )^{ \frac{1}{4} } \PA }_{1} 
    & \leq & 
    \norm{\check{X} \PA}_{1} + 
    \norm{ \PB ( 1 - \hat{X} ) \SigmaPsi \check{X} \PA }_{1} 
    \nonumber \\ 
    & \leq & 
    \norm{\check{X} \PA}_{1} +  
    \norm{(I - \hat{X}) \SigmaPsi X} \norm{\check{X} \PA}_{1}. 
\end{eqnarray}
Finally we estimate the operator norm of $(I - \hat{X}) \SigmaPsi X$ using~\eqref{eq: thermal_bounded_CAR} and by exploiting the finite propagation speed of the spinor wave opeator, 
$\partial_{t}^{2} + L, L := - \varDelta_{\mathscr{S\!C}} + R/4 + m^{2} - M^{2}$ 
following the original idea due to~\cite[Prop. 1.1]{Cheeger_JDiffGeom_1982} 
and subsequently generalized in~\cite[Appendix]{Hollands_JMP_2018}: 
\begin{eqnarray}
    \norm{(I - \hat{X}) \SigmaPsi X}
    & = & 
    \norm{(I - \hat{X}) \tanh \frac{\beta H}{2} X}  
    \nonumber \\ 
    & \leq & 
    \norm{(I - \hat{X}) \tanh \frac{\beta |H|}{2} X}  
    \nonumber \\ 
    & = & 
    \frac{1}{\pi} \norm{(I - \hat{X}) \int_{0}^{\infty} \widetilde{\tanh} (s) \sin (s \sqrt{L}) \rd (s) X} 
    \nonumber \\ 
    & \leq & 
    \frac{1}{\pi} \norm{I - \hat{X}}_{\infty} \norm{X}_{\infty} 
    \int_{d - 2 \varepsilon}^{\infty} \left| \widetilde{\tanh} (s) \right| \rd s 
    \nonumber \\ 
    & = & 
    \frac{1}{\beta} \norm{I - \hat{X}}_{\infty} \norm{X}_{\infty} 
    \int_{d - 2 \varepsilon}^{\infty} \left| \csch \frac{\pi s}{\beta} \right| \rd s 
    \nonumber \\ 
    & = & 
    - \frac{1}{\pi} \norm{I - \hat{X}}_{\infty} \norm{X}_{\infty} 
    \ln \left| \tanh \left( \frac{\pi}{2} \frac{d - 2 \varepsilon}{\beta} \right) \right|.  
\end{eqnarray}
Here we have used  
$H \leq |H|$, 
monotonicity of $\tanh$, 
and 
Schr\"{o}dinger~\cite{Schroedinger_1932}-Lechnwerowicz~\cite{Lichnerowicz_1963} formula to deduce $H^{*} H = L$ 
in the intermediate steps. 
The Fourier sine transformation $\widetilde{\tanh} (s)$ of $\tanh (\beta H/2)$  
is with prefactor $1/\pi$ in our convention and $\norm{\cdot}_{\infty}$ is the suprimum norm. 
%
%
%
%
%
%
%
%
%
%
\section{Relative entanglement entropy of Klein-Gordon QFT}
In this section, we present a minimal review of the quantization of a classical, real-linear Klein-Gordon field on any finite-dimensional $D \geq 3$, ultrastatic spacetime $(\sM, g) = \R \times \sC, \mathrm{d}t^{2} -h)$ and 
refer~\cite{Baer_Ginoux_Springer_2012}
and the 
expository articles~\cite{Kay_Wald_PR_1991, Sanders_IJMPA_2013} 
for details. 
We let 
$(\ks := C_{\mathrm{c}} (\sC, \R) \oplus C_{\mathrm{c}} (\sC, \R), \sigma)$ 
be the space $C_{\mathrm{c}} (\sC, \R) \oplus C_{\mathrm{c}} (\sC, \R)$ of Cauchy data of the linear Klein-Gordon equation   equipped with symplectic form $\sigma$. 

The algebraic Weyl algebra $\sW (\ks)$ over the symplectic space $(\ks, \sigma)$ is generated by the symbols 
$ W (0) := \mathbbm{1} $ and $W (F)$, 
modulo the relations~\cite{Kastler_JMP_1965, Manuceau_Verbeure_CMP_1968} 
(see also~\cite[A.2]{Baer_Ginoux_Springer_2012} 
and the expository articles~\cite{Kay_Wald_PR_1991, Sanders_IJMPA_2013}):  
\begin{equation}
	W (f)^{*} = W (- f), 
	\quad 
	W (f) W (g) 
	= 
	\exp \fbracket{ - \frac{\ri}{2} \sigma (f, g) } W (f + g),  
	\quad 
	\forall f, g \in \ks. 
\end{equation}
We turn the $*$-algebra $\sW (\ks)$ into a $C^{*}$-algebra $\overline{\sW (\ks)}$, called the Weyl algebra, by taking the completion with respect to the topology induced by the unique $C^{*}$-norm on $\sW (\ks)$. 

A thermal state $\Phi$ is a $\R$-linear functional of $\overline{\sW (\ks)}$, given by~\cite{Manuceau_Verbeure_CMP_1968}  
(see also the expository articles~\cite{Kay_Wald_PR_1991, Sanders_IJMPA_2013}):  
\begin{equation}
	\Phi (W (f)) 
	:= 
	\exp\fbracket{ - \frac{\scalarProd{f}{f}_{\Phi}}{2} }, 
	\quad 
	\forall f \in \ks,  
\end{equation}
where $\scalarProd{\cdot}{\cdot}_{\Phi}$ is a real inner product on $\ks$ satisfying 
\begin{equation} \label{eq: relation_inner_product_sympletic_form}
	\frac{1}{2} | \sigma (f, g) | 
	\leq 
	\sqrt{ \scalarProd{f}{f}_{\Phi} } \, \sqrt{ \scalarProd{g}{g}_{\Phi} }, 
	\quad 
	\forall f, g \in \ks. 
\end{equation} 
This results a real Hilbert space $(\ks, \scalarProd{\cdot}{\cdot}_{\Phi})$ 
(after taking the Hilbert space completion in the norm induced by $\scalarProd{\cdot}{\cdot}_{\Phi}$). 
It is convenient to introduce the complexification $\kS$ of $\ks$ 
and the extension (with the convention of being antilinear in the first argument) 
$\scalarProdDirac{}{}_{\Phi}$ of $\scalarProd{}{}_{\Phi}$ 
to describe the GNS representation of $\Phi$. 
Then we have the complex Hilbert space (after taking completion as before)
$(\kS, \scalarProdDirac{\cdot}{\cdot}_{\Phi})$ 
and the Riesz representation theorem implies that there exists 
a unique, bounded, selfadjoint operator $\Sigma$ on $\kS$ 
saturating~\eqref{eq: relation_inner_product_sympletic_form}: 
$\ri \sigma (f, g) / 2 = \scalarProdDirac{f}{\Sigma g}_{\Phi}$. 
Applying the GNS construction, we obtain the GNS triplet 
$(\pi_{\Phi}, \bkH, \Omega_{\Phi})$, 
where $\Omega_{\Phi} \in \C$ is the GNS-vector and 
$\pi_{\Phi}$ is the $*$-representation of $\sW (\kS)$ 
by bounded operators on the GNS-space $\bkH$ 
obtained from the symmetrized tensor product of 
the $1$-particle Hilbert space $\kH$, 
defined by factoring out $\kS$ by $\ker (I + \Sigma) / 2$ 
and taking the completion with respect to the GNS inner product  
\begin{equation}
	([F]|[G])_{\Phi} 
	:= 
	\frac{1}{2} \scalarProdDirac{F}{ (I + \Sigma) G}_{\Phi},  
	\quad 
	\forall F, G \in \kS. 
\end{equation}
Employing the GNS-representation, we define the local von Neumann algebra of observables as
\begin{equation}
    \sR (\sV) := \pi_{\Phi} \big( \overline{\sW (\kS)} \big)'', 
    \quad 
    \forall F \in \kS~|~\supp{F} \subset \sV, 
\end{equation} 
where $''$ is the double commutant, as before.

Entanglement and its relative entropy of a state of the algebra $\sR (\sC)$ 
is defined exactly the same way as those for the fermionic case $\sF (\sC)$. 
Without repetition, we state the main result of this section as

\begin{theorem} \label{thm: REE_Klein_Gordon_thermal}
    Let $\sC$ be a static Cauchy hypersurface in a geodesically complete, finite $D \geq 3$-dimensional ultrastatic spacetime 
    $(\sM = \R \times \sC, g = \mathrm{d} t^2 -h) $. 
    Then, for a thermal state $\Phi$ at an inverse temperature $\beta$ of the linear Klein-Gordon quantum field theory of mass $m$, 
    the relative entanglement entropy $E (\Phi)$ 
    between any open subsets $\sA, \sB \subset \sC$ 
    such that $d := \dist (\sA, \sB) \geq \delta > 0$, is bounded by 
    \begin{equation}
	    E (\Phi) 
	    \leq 
	    - \cst \ln \left| \tanh \frac{\pi}{2} \frac{d}{\beta} \right|. 
    \end{equation} 
    Here the positive constant $\cst$ is only $m, \delta$ 
    and the geometry within a $\delta$-neighborhood of $\sA$ dependant,  
    thus is independent of $d$ and $\beta$.
\end{theorem}
\textbf{Proof} 
The proof proceeds in parallel to that of our Theorem~\ref{thm: REE_Dirac_thermal}, 
when we note that $\Omega_{\Phi}$ is cyclic and separating~\cite[Thm. 4.8]{Strohmaier_CMP_2000} for $\sR (\sB')$ 
and $\Sigma := \tanh (\beta K/2)$ where 
$K : = \sqrt{ -\varDelta_{\sC} + m^{2}}$~\cite[Thm. 3.2]{Ogi_MathProcCamPhilosSoc_1991} 
(see also, e.g.,~\cite[Exam 5.3.2]{Bratteli_Springer_1997}).
So we consider the modular operator $\Delta_{\Phi, \sB'}$ 
for the pair $(\sR (\sB'), \Omega_{\Phi})$ 
and define the map 
\begin{equation}
    \Xi_{\Phi}^{\sA} : \sR (\sA) \to \bkH, 
    \quad A \mapsto \Xi_{\Phi}^{\sA} (A) 
    := 
    \Delta_{\Phi, \sB'}^{\frac{1}{4}} A \Omega_{\Phi}.
\end{equation}
As before, we are forced to estimate the $1$-nuclear norm of this operator.  
We can now implement the same strategy as in~\cite[Sec. V]{Hollands_JMP_2018}.
We defy the details as the steps will be quite analogous to those of Theorem~\ref{thm: REE_Dirac_thermal}. 
It is well-known that the time-reversal operator commutes 
with the Klein-Gordon Hamiltonian 
and due to its involutive nature, no doubling procedure is required. 
Then one can utilize it to construct the closed complex linear subspaces of $\kH$ 
and apply Propositions 5.2 and 5.3 with Theorems 3.11 and 3.5 in 
\cite{Lechner_Sanders_Axioms_2016} 
to deduce that the nuclear norm of $\Xi_{\Phi}^{\sA}$ 
is bounded by the trace-norm of  
$(P_{\sB'} - \Sigma_{\sB'}^{2})^{1/4} \! \upharpoonright \! \kS (\sA)$ 
where $\kS (\sA) := \kS \! \upharpoonright \! \sA$. 
The trace-norm estimation then follows from the 
positivity of the Klein-Gordon Hamiltonian $K$ 
(in contrast to the Dirac Hamiltonian) 
and exploiting the finite propagation speed of wave operator~\cite[Prop. 1.1]{Cheeger_JDiffGeom_1982}, 
as successfully contrived in~\cite[Prop. 4.3]{Lechner_Sanders_Axioms_2016}.

\section*{Acknowledgment}
The author is indebted to Ko Sanders for his guidance and pivotal comments.

\providecommand{\noopsort}[1]{}\providecommand{\singleletter}[1]{#1}%
\end{document}